Article

# Influence of the Process Parameters on the Quality and Efficiency of the Resistance Spot Welding Process of Advanced High-Strength Complex-Phase Steels

Gerardo Morales-Sánchez [1], Antonio Collazo [2,*] and Jesús Doval-Gandoy [1]

[1] APET Group, CINTECX, Campus As Lagoas—Marcosende, Universidade de Vigo, 36310 Vigo, Spain; gmorales@uvigo.es (G.M.-S.); jdoval@uvigo.es (J.D.-G.)
[2] ENCOMAT Group, CINTECX, Campus As Lagoas—Marcosende, Universidade de Vigo, 36310 Vigo, Spain
* Correspondence: acollazo@uvigo.es

**Abstract:** In this study, the effects of electrical characteristics of an inverter combined with main welding parameters on the resistance spot weldability of advanced high strength steels (AHSS) CP1000 is investigated. The main welding parameters, current and time, were varied. The effects on the geometry and microstructure of the weld spot, the diameter of the weld pad, the hardness, the shear strength, and the efficiency of the process were studied, and the results were compared for two switching frequencies of the medium frequency converter. Furthermore, the weldability lobes were obtained as a function of the shear strength, for both frequencies. This work shows that the quality of the welding, in the established terms, is better when a lower frequency is used, even though the parameterization of the welding equipment can be easier for higher frequencies.

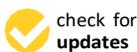

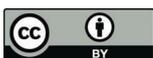

## 1. Introduction

The continuous evolution of industry, especially of the automotive industry, has led to an interest in the continuous development of state-of-the-art materials, and in particular that of advanced high-strength steels. These steel types provide notable benefits, allowing the manufacture of lighter vehicles, with lower fuel consumption, lower emissions and greater autonomy, in the case of electric vehicles, while maintaining high safety standards. In order to reduce costs in the final product, the aim is to reduce the thickness while maintaining the formability of the material to facilitate stamping and die-cutting operations. These aspects are achieved through an adequate compromise between strength and ductility [1–5].

The industrial use of AHSS, which include dual phase steels, complex phase steels, transformation-induced plasticity steels and martensitic steels, is conditioned by the adaptability of the processes and the conventional equipment to these new materials. From this point of view, it should be noted that the RSW process is one of the most widely used in the automotive world due to its low cost and high cycle speed. Each car can include about of around 4500 resistance welding spots, making it the main joining method in the automotive industry [6–9].

The resistance welding process has been studied by many authors [10–12]. It is a process used to join metals thanks to the heat generated by the resistance offered by the pieces to be joined to the flow of the electric current and the application of pressure through the electrodes or caps. The size and geometry of the electrode are important factors for the welding quality, and the manufacturers' recommendations vary depending on the material [13]. The heat generated is based on Joule's law: $Q = R \cdot I^2 \cdot t$ where, Q is the heat, I the welding current, t the welding time and R the resistance. The latter can be decomposed into seven partial resistances, which depend on various parameters such as the material, the pressure or cooling [6,14]. The higher the resistance, the higher the temperature reached at that point and therefore the larger the size of the melting zone.

The combination of phases with very different thermomechanical behaviors gives rise to a high microstructural complexity of the CP, bainite, ferrite and martensite [15], so the failure mechanisms of the RSW-welded joints of all these steels have not been studied in depth [16]. In addition, due to their high carbon equivalent content, problems related to the formation of martensite in the weld bead can occur during welding of these steels, therefore the search for the weldability lobe is of clear interest. In order to have an efficient industrial welding process it is necessary to properly choose the welding parameters. The main parameters that influence the quality of the welding are: the welding current, I; the welding time, t; and the welding force, F, which is inversely proportional to the resistance [6,12]. Other important parameters are: the diameter of the active face of the electrode; the material to be welded and its coating; the type of power supply; the design of the machine and many other factors that make this apparently simple process one of great complexity [10,11,17]. The weldability lobe or weldability range defines the tolerances to produce quality welds, and facilitates the parameterization of welding equipment in the industry.

Recent publications [18–20] have sought to optimize the welding parameters of some AHSS based on the metallographic study, the tensile strength and the type of failure, concluding that the key characteristics are controlled by phase transformations. In general the increase of welding current gives rise to a greater force to the shear, and there is a great variation in hardness along the cross section in the steels of high resistance attributed to rapid heating and cooling during welding that generate phase transformation. AHSS undergo complex microstructural transformations during the RSW process, the failure mode is related to hardness and microstructure in the fusion zone (FZ) and in the heat affected zone (HAZ). Regarding the AHSS complex phase, the predominant failure is the interface failure when there is an absence of Boron [21,22]. The tendency to interfacial failure of AHSSs means that a large button diameter size does not always result in button failure. A study of the cycle of the welding process can improve the mechanical characteristics. Therefore, the need to define new quality criteria is established [8].

The RSW equipment varies depending on the final application for which it has been designed, being of great importance its correct design [23], structure and dimensioning to optimize the functionality, repeatability, quality and efficiency of the process. In RSW equipment, the power blocks have progressed in their technology, going from the most common single-phase power conversion stage at mains frequency to the power converter of medium frequency to direct current (MFDC) that work at frequencies in the kHz range. Different studies [24–27], have demonstrated the advantages of MFDC technology over alternating current (AC). Among others, the weight reduction in transformers by using MFDC technology with a frequency of 1000 Hz or higher, allows this element to be placed in a robotized gripper, avoiding an extra cost in the acquisition of a robot of greater weight and volume. In this way, the robotization and automation of welding processes is facilitated, helping their incorporation into assembly lines [12,23,28] and therefore achieving a reduction in production costs. The high currents required in the RSW process give rise to high temperatures, such that the greatest energy loss is in the transformer and in the interconnection and electrodes of the secondary circuit, which makes necessary an efficient cooling system. One way to increase power density while reducing cooling requirements is to reduce power loss in these elements.

Regarding the switching frequencies in MFDC equipment, in [29] it is established that the switching efficiency improves as an inverse function of the frequency. Increasing the switching frequency allows the use of smaller and lighter transformers [12]. The type of transformer and its design are fundamental factors in the efficiency of the welding process. The most significant losses are due to the rectifier located between the secondary of the transformer and the electrodes. An improved distribution of the windings results in a decrease in saturation and consumption [30–33].

At present, industrial processes are highly automated and penetrated by robotic systems, in this sense the use of RSW equipment based on MFDC is justified. In addition, the frequency of work of the MFDCs has been gradually increased in recent years in order to have lighter, less bulky, more precise and faster welding equipment.

Despite studies conducted in the past in relation to these technologies, it has not carried out a detailed study on how it affects the switching frequency of the inverter to the quality of the welding of AHSS, the lobes of weldability, the parameterization of the welding process and to the electrical performance of the assembly.

This article presents a study of the influence of the inverter working at frequencies of 1 kHz and 7 kHz on the quality and efficiency of the RSW process in CP1000 AHSS. In order to verify the performance of the process, two quality metrics are established: the efficiency of the electronics conversion stage and the quality of the welded joint.

In Section 2 the materials, the equipment used and the methodology are described, in Section 3 the results are presented and the discussion of these is carried out and finally in Section 4 the conclusions of the article are presented.

## 2. Materials, Equipment and Methods

### 2.1. Base Material

For this study, it has been employed a galvanized steel complex phase CP1000 G10/10, used in the automotive industry. The coating thickness of the zinc layer is 10–13 µm on each side. Table 1 shows the chemical composition, provided by the manufacturer, Arcelor Mittal [34].

**Table 1.** Chemical composition of CP1000 G10/10 steel.

| Element | C    | Mn  | S     | P     | V     | Al  | Si   | Cr   |
|---------|------|-----|-------|-------|-------|-----|------|------|
| wt%     | 0.17 | 2.0 | 0.010 | 0.030 | 0.030 | 0.1 | 0.35 | 0.60 |

### 2.2. Equipment

This study focuses on RSW based on MFDC technology. The main objective is to compare the quality of the welded joint and the efficiency of the welding equipment for two different working frequencies, 1 kHz and 7 kHz. Figure 1 shows the scheme of the welding equipment built to develop this study. All resistance welding points were made using a Safco CSI 450A (Safco Systems, Cesano Boscone, Italy), composed of a three-phase rectifier and an inverter operating in a wide range of switching frequencies between 1 kHz and 7 kHz. At the inverter output there are two Elesa transformers (Elesa Trafocore, Guipuzcoa, Spain) in parallel and the corresponding rectification stage. One of the transformers is designed to work at frequencies around 1 kHz and the other to work at frequencies around 7 kHz, and the output of the transformers is taken to the same secondary block of interconnections and electrodes, where the welding point is made. In Figure 2a it is shown a general image of the equipment used for welding, Figure 2b shows the physical arrangement of the transformers in the equipment.

The voltage-current pairs of the mains input, the primary of the transformer and the output of the transformer-rectifier were measured. The efficiency of the welding equipment is calculated as the ratio of useful energy that comes out of the output rectifier device to the total energy drained from the three-phase electrical grid. The voltages ($V_g$, $V_{pri}$, $V_{sec}$) were measured with voltage probes connected to an oscilloscope that allows the data to be stored in .csv format for later treatment. The current at the mains input ($I_g$) and at the primary ($I_{pri}$) are measured with rogowski current probes connected to the oscilloscope for further treatment. The secondary current ($I_{sec}$) is captured with a Rogowski coil that, with its conditioner, allows it to be connected to an oscilloscope.

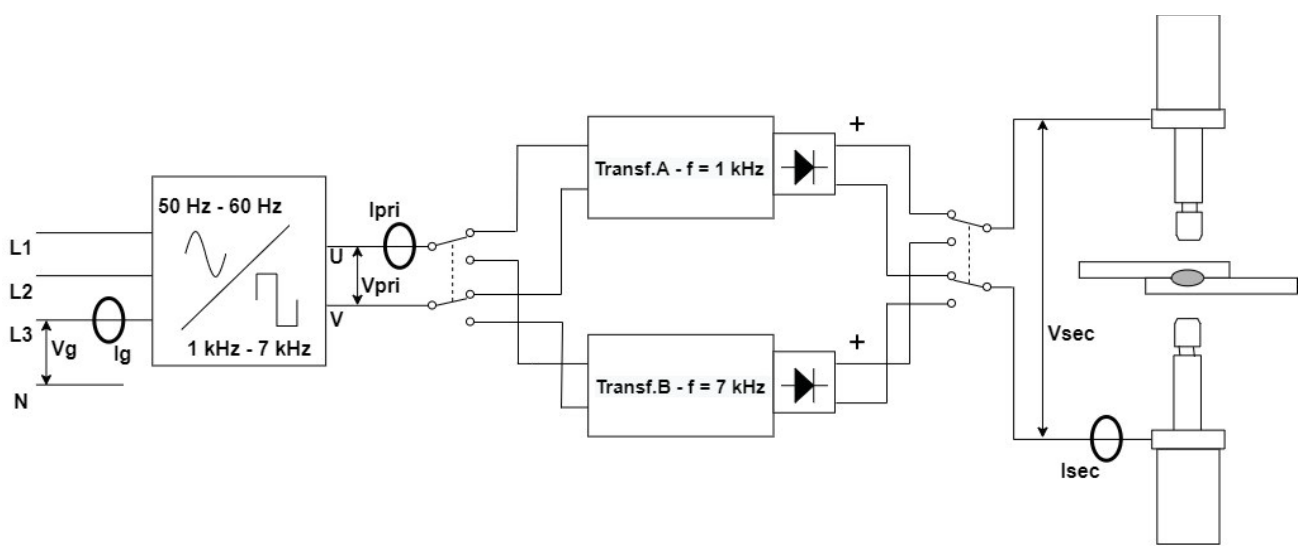

**Figure 1.** Schematic of the welding equipment used in the experiments.

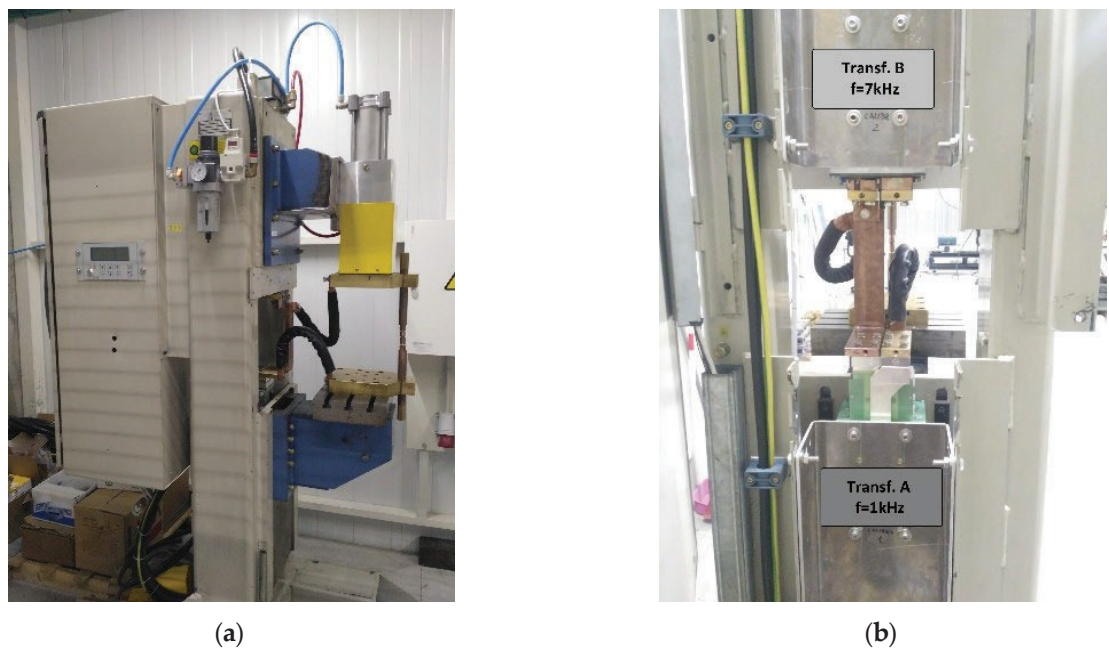

(**a**)　　　　　　　　　　　　　　　　　(**b**)

**Figure 2.** Welding equipment: (**a**) side view; (**b**) the arrangement of the transformers.

### 2.3. Tests and Methods

The welding current is passed in a single pulse through the B-16 electrodes made of Cu–Cr–Zr alloy and with an active face of diameter of 6 mm, according to the ISO-5821 standard [35]. In order to minimize the influence of the electrode wear, a procedure is established to review their condition after carrying out a minimum number of welds, replacing them with new electrodes. This allows the comparative study to be focused on, thus being able to identify the influencing factor for each of the aspects studied. After each replacement of the electrodes, a minimum of 10 welds are made, which allow them to be adapted before making the welding points of the study.

To carry out this study, a total of 556 AHSS CP1000 G10/10 steel specimens were welded, Table 2. To obtain the weldability lobe, the welding force F of 470 daN was set; the frequency was varied between 1 and 7 kHz. Welding tests were carried out with various values of intermediate frequencies between 1 kHz and 7 kHz. The results of these tests are aligned with that shown in the figures of the article; the trend line is maintained. There are no maximum or minimum values for intermediate frequency values that break with the trend line. A variety of welding times were used, 170, 210, 220, 315, 420 and 525 ms, and the current was varied in the range (5.5–14 kA). In addition, to justify the $F_m$ as a quality criterion, welds were carried out at a force F of 572 daN, and a welding time t of 260 ms, for welding currents of 10 to 11 kA.

Table 2. Specimens tested.

| Freq. (Hz) | Specimens | Ensayos | | | | | | Total Tests |
|---|---|---|---|---|---|---|---|---|
| | | η | Fm | Φ | HV01 | Macro | SEM | |
| 1000 | 285 | 244 | 180 | 221 | 26 | 27 | 7 | 705 |
| 7000 | 271 | 237 | 159 | 198 | 26 | 26 | 9 | 655 |
| Total | 556 | 481 | 339 | 419 | 52 | 53 | 16 | 1360 |

Efficiencies was obtained from the test of 481 welded specimens; the diameter of 419 weld points were measured; 339 shear strength tests were performed; 52 hardness tests; 53 welding points were analyzed macrographically and 16 through SEM. A total of 1360 tests were carried out on the welded specimens.

The welding force was set at a force F of 470 daN, a value that is estimated at the center of the weldability lobe at constant time, based on previously carried out experimental work and references [3,36,37]. The time and current values were selected following the procedures established in the standards for obtaining the weldability lobe, ISO 14327 [38], initial criteria ISO 14373 [39] and for obtaining quality welds, ISO 18278-2 [37]. In this way, through experimental tests, starting from a time of 420 ms that was considered centered on the lobe and as has been demonstrated, three other time values were established as indicated by the ISO 14327 standard [38]. The current steps were given according to the procedure established in ISO 18278-2 [37] until reaching the lower limits (Minimum diameter) and upper limits (Projections).

The frequency values were established as the most extreme values that commercial equipment allows and which represent the two types of equipment that are present in the industry, the static pedestal machine (1 kHz) and the robotic clamp (7 kHz). The tests carried out include those necessary for the establishment of the weldability lobe according to ISO-14327 [38]. All significant welding points were repeated a minimum of three times under identical conditions, from these values the mean and deviation that will be used in the representation and treatment were obtained.

The test specimens made by overlapping welds on two identical samples of the study material were carried out on two types of coupons based on the tests for which the welding has been intended: a "Point coupon", with dimensions greater than three times the diameter of the welding point, which were intended for microhardness tests, macrographic study and analysis by scanning electron microscopy (SEM); and a "Shear coupon", with dimensions according to ISO-14273 [40] or equivalent, which were intended for evaluate the shear strength and identification of the type of failure.

The welded joint strength was evaluated from shear test using a Universal Testing Machine (Shimadzu AG-I, Shimadzu Corporation, Kyoto, Japan) with a capacity of 250 kN. The tests were performed using a strain rate of 10 mm min$^{-1}$, at room temperature. This test has been chosen because it is considered more reproducible and reliable in the resulting data compared to the chisel test. It has been verified [41] that, depending on the test carried out (shear or chisel), the type of failure of the weld will be one or the other. It has been verified that the chisel test shows a greater number button failures (BF), compared to the predominant type of interface failure (IF) in the shear strength. It has been verified that the diameters of the welding lentil measured in specimens tested with a chisel have an average diameter greater than those measured in specimens tested by shear, so the use of the data obtained in the shear strength is more restrictive. The type of failure and how to measure the diameter is established in ISO-14329 [42].

The rest of the specimens were used for Vickers hardness, microscopy and SEM tests. The specimens called "Point coupon" were cold cut with the Struers Secotom-10 cutter (Struers, Rødovre Denmark), and hot mounted with the Metkon Ecopress 102 equipment. The samples were mechanically ground with successive grades of SiC papers to 1200 grade, after which they were polished with diamond paste of 1 µm grain size to obtain a mirror-like appearance with the Buehler Metaserv Motopol 8 equipment (Buehler Metaser, Buehler, IL, USA). Finally, the samples were rinsed using ethanol and distilled water, and were air-dried. The hardness test HV01 (Vickers 100 gr) was carried out according to ISO-14271 [43], with the Emcotest DuraScan equipment (Traunstein, Germany). The microstructure was revealed by etching with nital (2% nitric acid in ethanol, volumetric), all the chemical reagents were supplied by Sigma-Aldrich (Merck KGaA, Darmstadt, Germany). The surface morphology was studied by optical microscopy and SEM. The equipment employed was an optical microscopy Olympus GX51 equipment (Tokio, Japan) and an electron microscope Electroscan JEOL model JSM-6510. The chemical composition of the phases was determined by energy dispersive X-ray spectroscopy (EDS) with the Link ISIS 300 X-ray detector. The acceleration voltage was 20 kV, both are from OXFORD instrument (Abingdon, England).

*2.4. Weldability Lobe Criteria*

To establish the initial values of the weldability lobe, the recommendations of the ISO-14327 [38], ISO-18278-2 [37] and AWS C1.1 [17] standards were followed. As a welding quality requirement, the tensile shear strength $F_m$ [kN] was adopted from among those proposed in ISO-14373 [39]. The value of $F_m$ used was calculated according to [3], $F_{mok}$ = 15.83 kN. A weldability lobe will be made for each of the frequencies under study, 1 kHz and 7 kHz.

*2.5. Quality Criteria*

As previously mentioned, two quality criteria were used: the electrical efficiency of the welding equipment and the quality of the welded joint. The first is obtained through efficiency η, calculated as the ratio of output energy and input energy. The second is obtained by means of the shear strength force $F_m$ (kN), although the classic measurement is also presented, the diameter of the weld point φ [mm]. Complementary to these parameters, aspects such as hardness analysis, macrographs and SEM are also presented, analyzing parameters in this way the welded joints from a metallurgical and electrical point of view, the use of $F_m$ as a quality criterion is supported by ISO-14373 standards [39] and by authors [8,19,36] who state that the criteria should be re-evaluated.

There are several studies that, from an electrical point of view, use the energy efficiency to establish comparisons and conclusions in the resistance spot welding process [25,27,30]. Despite the fact that there are several published studies, none of them relate the electrical aspects with the quality of the weld, making an in-depth study of its mechanical characteristics and defects. For the efficiency calculation, energy will be used, since our signal is not periodic. From the real-time signals of V and I that have been captured and processed, the energy can be calculated using the formula:

$$E = \int^r V(t) * I(t) dt \tag{1}$$

Applying (1) to the V-I pair of the main input, and to the secondary pair, we obtain:

$$E_g = 3 * \int^r V_g(t) * I_g(t) dt \tag{2}$$

$$E_{sec} = \int^r V_{sec}(t) * I_{sec}(t) dt \tag{3}$$

Equation (4) is used to calculate the efficiency of the welding process at two different frequencies, 1 kHz and 7 kHz.

$$\eta = \frac{E_{sec}}{E_g} \quad (4)$$

## 3. Results and Discussions

### 3.1. Study of the Base Material

To determine the strength and ductility characteristics of the base material, uniaxial tensile tests were performed according to ISO-6892-1 [44] on two specimens in the longitudinal direction of the rolling and two specimens in the transverse direction. Table 3 shows the yield stress values ($R_e$), tensile strength ($R_m$), percentage elongation after break (A%) and percentage reduction in thickness after break (Z%). All values were within the ranges established by the manufacturer.

**Table 3.** Tensile test. Characterization CP1000 G10/10.

| Mechanical Characteristic | Longitudinal_1 | Longitudinal_2 | Transversal_1 | Transversal_2 |
| --- | --- | --- | --- | --- |
| $R_e$ (MPa) | 945.7 | 904.7 | 861.8 | 912.8 |
| $R_m$ (MPa) | 1023.9 | 997.8 | 996.5 | 997.8 |
| A% | 11 | 11.78 | 8.65 | 8.8 |
| Z% | 9.4 | 13.9 | 20.6 | 22.8 |

Longitudinal_1 and Transversal_1 coupon fractography was performed, obtaining similar results for both coupons. The samples suffer a high necking from 1.8 mm of the nominal thickness to about 0.8 mm.

Figure 3a shows the fracture lip in the central crack. The presence of micro-voids in both coupons identifies a clear ductile character on a microscopic level. A large number of nonequiaxed dimples of heterogeneous size, even though two families are predominant, with 1–2 μm and 4–5 μm diameters. This size distribution can be qualitatively correlated with the particle size of the second phases (ferrite and martensite). The presence of elongated dimples is consistent with shear loading because, as is well known, dimple morphology is influenced by loading conditions, as can be seen in Figure 3b [45].

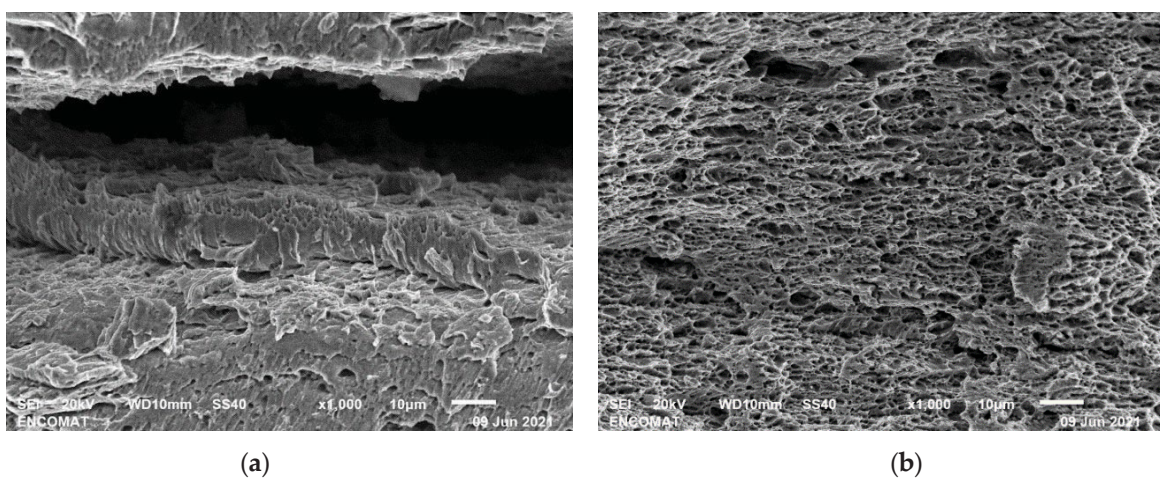

(a) (b)

**Figure 3.** Central crack fracture: (**a**) the fracture lip; (**b**) the presence of micro-voids.

### 3.2. Weldability Lobes

Weldability lobes are a useful tool for finding the optimal welding parameters. In this study, two lobes were developed, one at f = 1000 Hz, Figure 4a, and the other at f = 7000 Hz, Figure 4b. The elaboration of these lobes has been carried out based on the reference standards ISO-18278-2 [37] and ISO 14327 [38]. The key indicators we obtained from the 556 welding points and 1360 tests. These key indicators consist of minimum diameter, projection diameter, and optimal welding, which were marked, respectively, with diamond, triangles and squares. The minimum value is defined by $D_{min} = 3.5 * \sqrt{t}$, where $t = plate\ thickness$ [mm], in this case a diameter $D_{min}$ of 4.7 mm, and the maximum value is limited by weld projections. The optimum value has been determined based on the maximum shear force supported by the joint $F_m$ [N], a quality requirement commonly established in the industry and supported by the ISO-14373 standard [39].

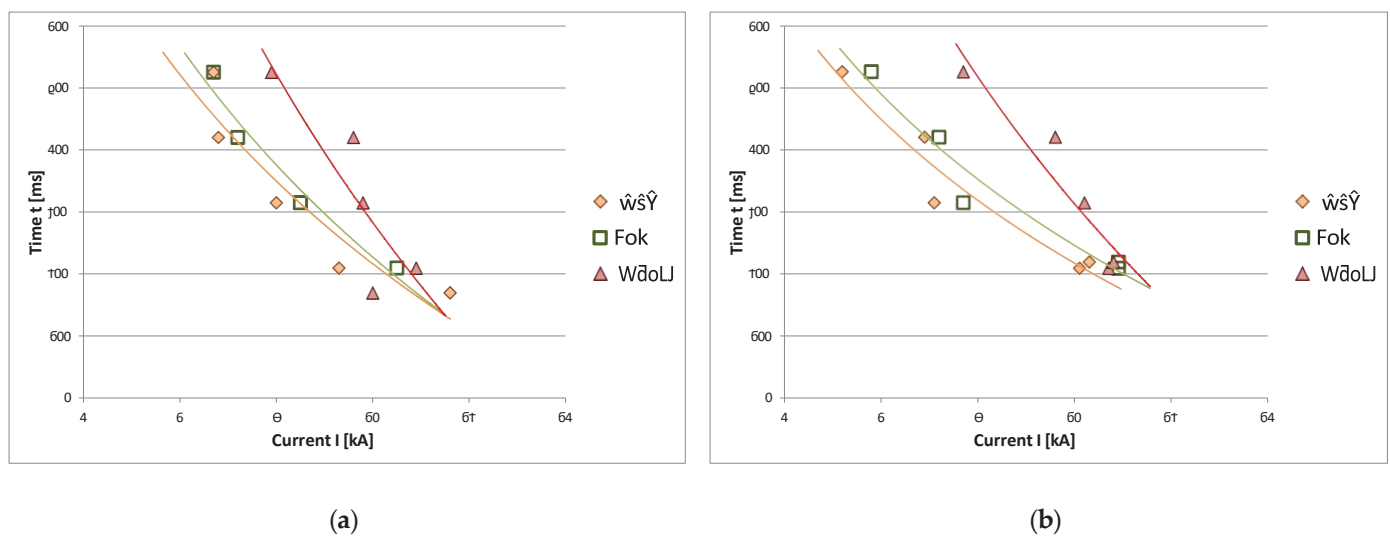

(**a**)　　　　　　　　　　　　　　　　　　(**b**)

**Figure 4.** Key indicators for the lobes obtained after 556 welding points and 1360 tests including minimum diameter (diamond), projection diameter (triangle) and optimal welding (square). The lines show the welding lobes following ISO 18278-2 [37] and ISO 14327 [38] AHSS CP1000 G10/10 steel weldability lobes: (**a**) f = 1000 Hz; (**b**) f = 7000 Hz.

Table 4 shows the electric current values in kA of the significant points for each timeline and for each frequency, and also shows the working range in which quality welds can be performed. $ID_{min}$ indicates the current values immediately lower than those necessary to reach the minimum diameter of the weld spot. $IF_{ok}$ is the estimated value of the curves to obtain a valid $F_{ok}$ value. $I_{Proy}$ shows the current value from which there are projections, it is established when half plus one of the welding points made with that current intensity result in projections.

**Table 4.** Weldability lobes.

| | Frequency | | | | | | | |
|---|---|---|---|---|---|---|---|---|
| | f = 1000 Hz | | | | f = 7000 Hz | | | |
| Time (ms) | $ID_{min}$ (kA) | $IF_{ok}$ (kA) | $I_{Proy}$ (kA) | Range (kA) | $ID_{min}$ (kA) | $IF_{ok}$ (kA) | $I_{Proy}$ (kA) | Range (kA) |
| 170 | 11.6 | - | 10 | - | - | - | - | - |
| 210 | 9.3 | 10.5 | 10.9 | 1.6 | 10.1 | 10.9 | 10.7 | 0.6 |
| 220 | - | - | - | - | 10.3 | 10.9 | 10.8 | 0.5 |
| 315 | 8 | 8.5 | 9.8 | 1.8 | 7.1 | 7.7 | 10.2 | 3.1 |
| 420 | 6.8 | 7.2 | 9.6 | 2.8 | 6.9 | 7.2 | 9.6 | 2.7 |
| 525 | 6.7 | 6.7 | 7.9 | 1.2 | 5.2 | 5.8 | 7.7 | 2.5 |

As shown in Figure 4, in general the current range was greater for the 7000 Hz frequency and the weldability lobe is shifted to the left, that is, towards smaller current values. All these indicators show that the parameterization of a welding equipment that works at a switching frequency of 7 kHz is, initially, simpler.

### 3.3. Geometry and Microstructure of the Weld Point

Figure 5 shows the macrographs made to significant specimens in the areas of diameter of the weld point below the minimum, $D_{min}$ (a), and in the areas of projections, $D_{Proy}$ (b), for the two test frequencies considered.

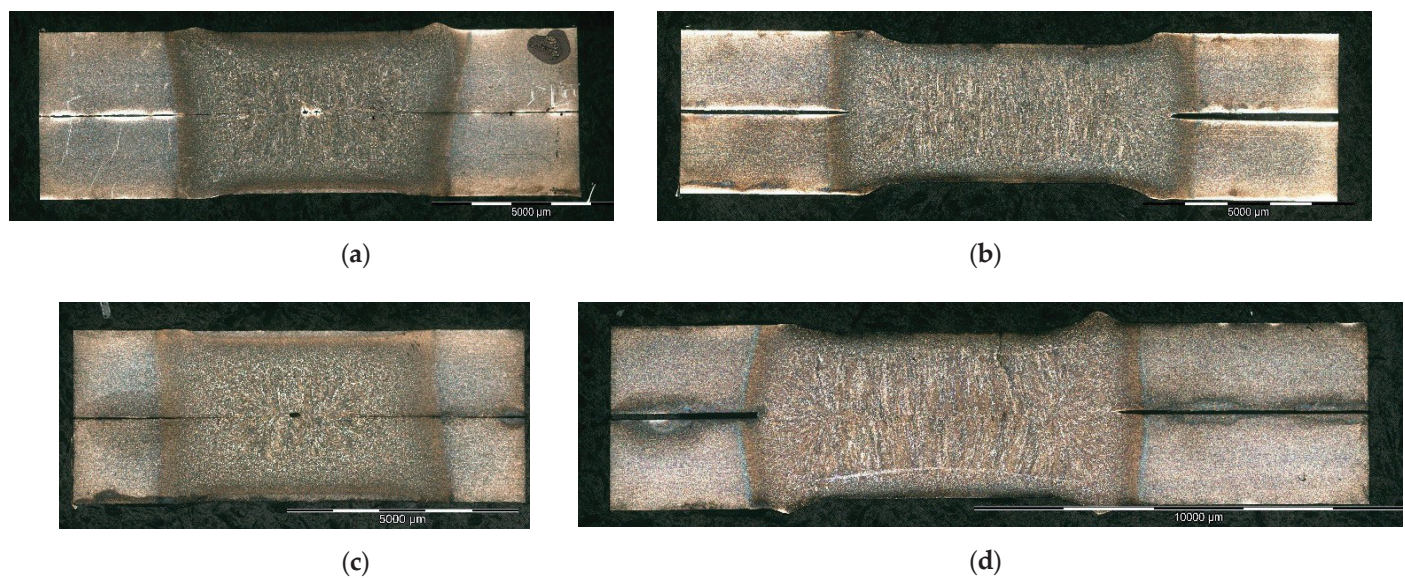

**Figure 5.** Macrography for f = 1000 Hz: (**a**) $D_{min}$ for f = 1 kHz; (**b**) $D_{Proy}$ for f = 1 kHz; (**c**) $D_{min}$ for f = 7. kHz; (**d**) $D_{Proy}$ for f = 7 kHz.

It can be seen that the shape of the welding point is more uniform, ellipsoidal, for switching frequency of 7 kHz in the case of $D_{min}$ values, and it becomes more uniform for switching frequency of 1 kHz for $D_{Proy}$ values. Specimens corresponding to $D_{min}$ values have internal pores in the center of the weld point. For the case of $D_{Proy}$ at 7 kHz a wide crack can be observed from the surface of the upper face.

Table 5 includes the mechanical characteristics for the significant values. $D_{min}$ represents the points immediately below the minimum diameter, $F_{ok}$ is the first value that reaches the desired $F_m$ and $D_{Proy}$ is the first value wherein the projections are the majority. From this table it can be concluded that, although the repeatability at $D_{min}$ values is better for 1 kHz, this trend reverses as the welding current is increased until it reaches $F_{ok}$ values for which repeatability is better for 7 kHz. The decrease in hardness of the fusion zone is surprising in the cases of welds carried out at a switching frequency of 7 kHz, which seems to indicate that the switching frequency modifies the output energy enough to carry out a microstructural variation.

**Table 5.** Mechanical characteristics. Weldability lobe.

|  | Frequency | | | | | |
| --- | --- | --- | --- | --- | --- | --- |
|  | f = 1000 Hz | | | f = 7000 Hz | | |
| **Values** | $D_{min}$ | $F_{ok}$ | $D_{Proy}$ | $D_{min}$ | $F_{ok}$ | $D_{Proy}$ |
| Diameter φ(mm) | 4.4 | 5.1 | 6.8 | 4.5 | 5.8 | 8.0 |
| Standard desviation | 0.1 | 0.3 | 0.6 | 0.1 | 0.3 | 0.1 |
| Tensile Shear Force $F_m$ (kN) | 11.1 | 15.8 | 22.4 | 11.9 | 15.9 | 32.1 |
| Standard desviation | 0.5 | 1.8 | 9.3 | 1.3 | 1.3 | 1.1 |
| Hardness FZ (HV01) | 424.1 | 413.8 | 426.5 | 366.9 | 303.3 | 400.1 |
| Standard desviation | 44.4 | 62.6 | 35.0 | 85.0 | 16.7 | 43.4 |

Figure 6 shows a detail of the welded joint structure for the $D_{min}$ value corresponding to switching frequency of 1 kHz, identifying the base metal regions (BM), heat affected zone (HAZ), and fusion zone (FZ). The lack of metallurgical continuity in the area of the HAZ and the beginning of the molten area is evident. The spectrum of energy dispersive X-ray made in these regions reflects the presence of aluminum and zinc along the interface. In the central region of the fusion zone in Figure 6b, a pore of 0.54 mm diameter can also be observed, which represents 15% of the diameter of the weld point, significantly affecting the strength of the joint. In addition, the geometry of the pore is clearly irregular, ending in cracks that can propagate due to fatigue when the material is subjected to cyclical stresses.

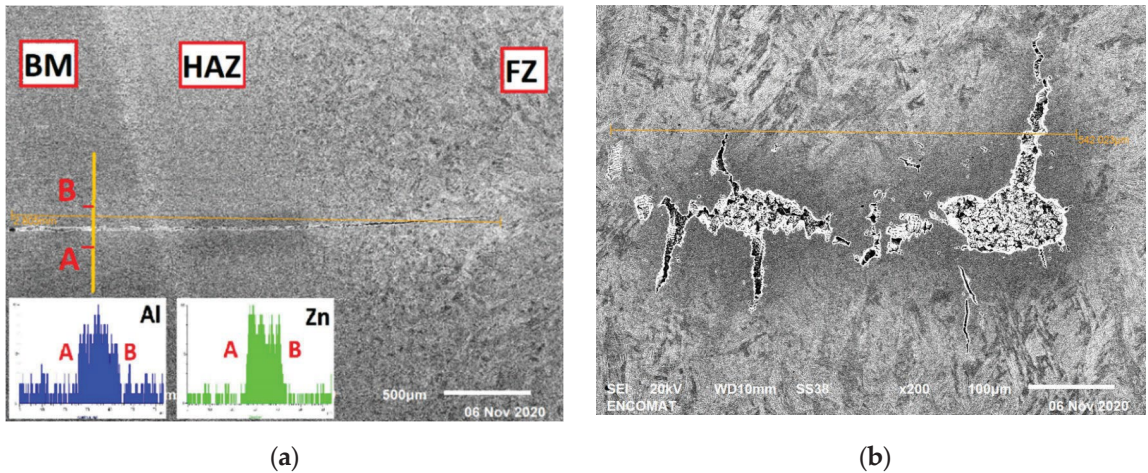

**Figure 6.** SEM of $D_{min}$ for f = 1000 Hz: (**a**) the nterface; (**b**) detail of the central defect in the FZ.

Figure 7 shows the SEM images of the HAZ and FZ corresponding to the $D_{min}$ value at 1 kHz, with a mainly martensitic structure [36], in the case of the FZ the columnar structure can be seen, no differences were observed with respect to 7 kHz.

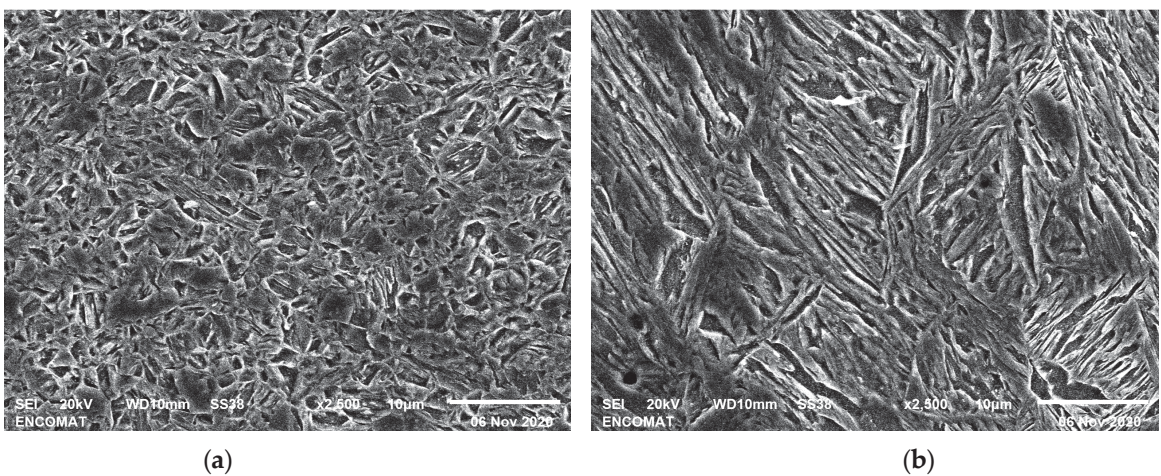

**Figure 7.** SEM of $D_{min}$ for f = 1000 Hz: (**a**) heat affected zone (HAZ); (**b**) fusion zone (FZ).

In Figure 8 the SEM image is presented for the $D_{Proy}$ value corresponding to switching frequency of 1 kHz, in Figure 8a the interface can be seen, as in the case of the $D_{min}$ value there is incomplete expulsion of Zn, but in this case of diameter of 2.09 mm, that is, 25.5% less, which confirms that, when moving to the right in the weldability lobe, towards expulsion values, the energy supplied to carry out the welding is increased, which helps the expulsion of the Zn. In this case, the incomplete interface is in the HAZ without entering the FZ. Figure 8b shows the SEM image of the BM for this value.

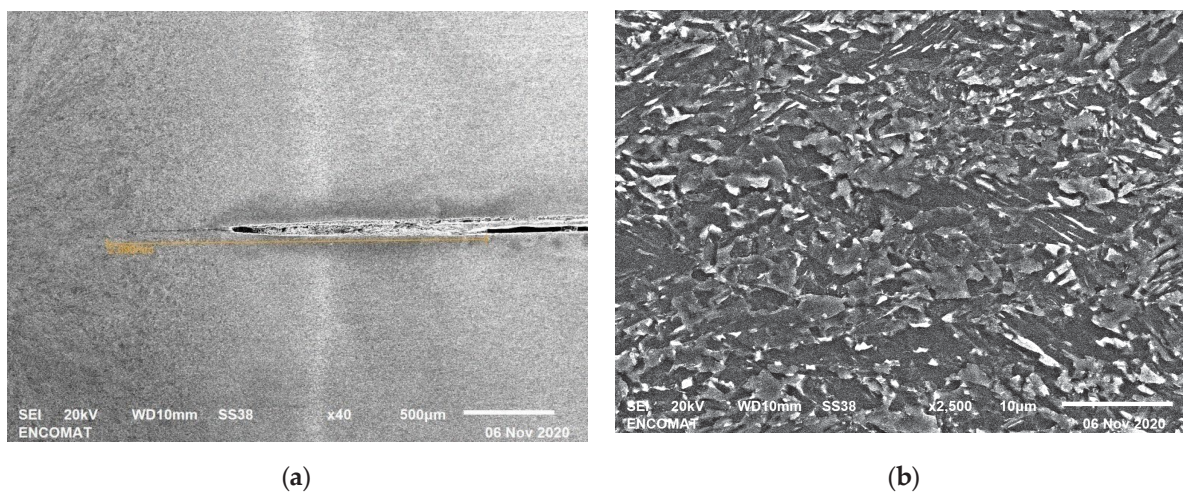

**Figure 8.** SEM of $D_{Proy}$ for f = 1000 Hz: (**a**) the interface; (**b**) the base metal.

Figure 9 depicts the interface for the case $D_{min}$ at a frequency of 7 kHz, the measurement of the Zn crown is clearly less than in the case of 1 kHz, since it has gone from a diameter of 2.805 mm to diameter of 1.2 mm of incomplete Zn ejection for 7 kHz. This aspect is a clear advantage, showing that the increase in frequency represents a considerable improvement in the ejection capacity, seeing this ring reduced to less than half. In any case, the presence of Zn is still important, which can cause repeatability problems, as can be seen in Figure 9b [46–48]. This specimen, $D_{min}$ at a switching frequency of 7 kHz, has a central pore of about diameter of 0.24 mm, less than half the size corresponding to the value Dmin at a switching frequency of 1 kHz of diameter of 0.542 mm, as it is shown in Figure 7b.

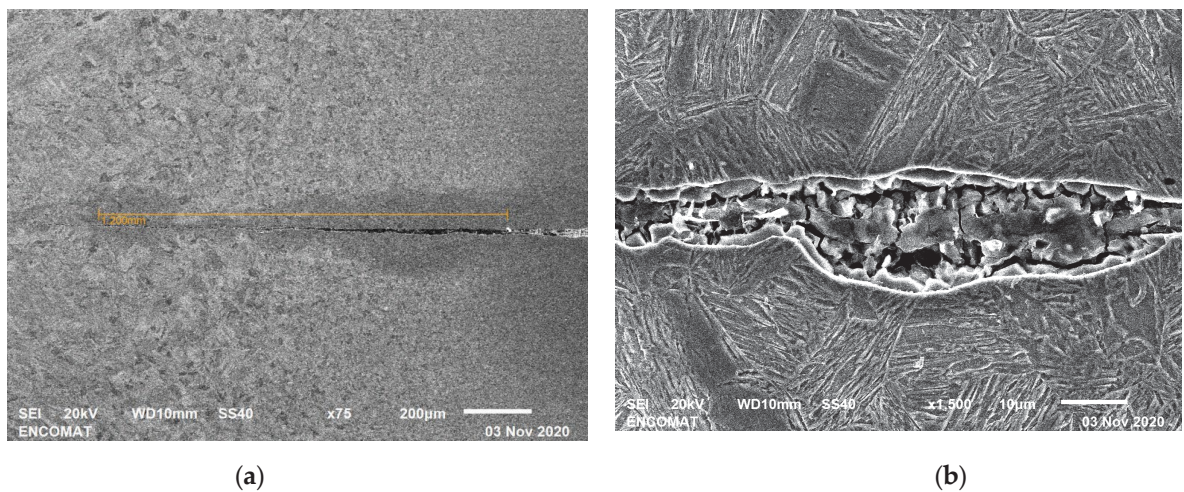

**Figure 9.** SEM of $D_{min}$ for f = 7000 Hz: (**a**) the interface (**b**) detail of Zn in the interface.

According to the manufacturer, the zinc layer present in the welded samples is between 10–13 µm, but it has been possible to measure a zinc layer with a thickness of 17 µm, this variation may be enough so that the parameters optimized for the factory values are not optimal in this case, causing incomplete ejection or a lower quality weld [9,48,49].

In Figure 10 is depicted the beginning of the FZ has the presence of Zn, this zone is not separated to the naked eye, but it has an incomplete ejection zone of about diameter of 0.278 mm, which confirms that, when advancing to the right in the weldability lobe, the expulsion of Zn is improved. Comparing with the equivalent DProy value at 1 kHz, it has gone from a diameter of 2.09 mm to a diameter of 0.278 mm for 7 kHz, which represents a decrease of 86.5%, which confirms that the increase in frequency improves the ejection of the Zn and therefore the repeatability of the welds. In any case, there is still the presence of Zn in the interphase as can be seen in Figure 10b at the beginning of the FZ.

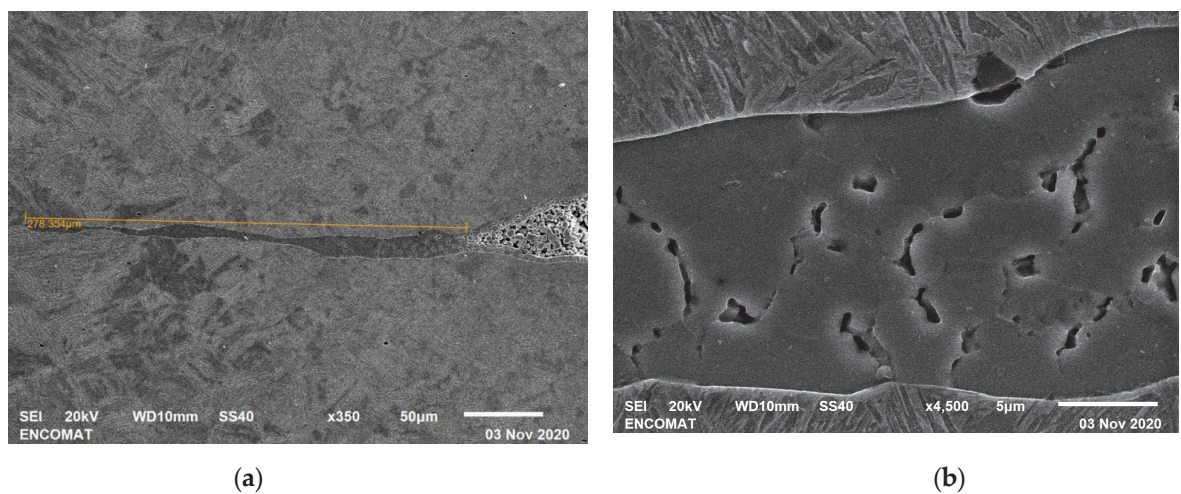

(**a**)  (**b**)

**Figure 10.** SEM of $D_{Proy}$ for f = 7000 Hz: (**a**) the interface (**b**) detail of Zn in the interface.

*3.4. Weld Spot Diameter*

The diameter of the weld spot is one of the most used quality criteria in the industry. Figure 11 shows the evolution of the size of the welding spot as a function of the welding current for both frequencies, keeping welding time and welding force constant. The points corresponding to the means of the diameters, the deviation for each point and the trend lines are displayed. The vertical lines correspond to the interpolated current values to establish the functionality and quality of the welded joints. They were obtained from the diameter–current graph, considering the trend lines of the point clouds and the cut-off value for $D_{min}$, $F_{ok}$ and $D_{ok}$. $D_{min}$ is the current value for which, of the series of welds carried out, half plus one of the measured diameters is less than the minimum diameter; $D_{proy}$ corresponds to the current value, for which the series of welded specimens has projections in most of them; $D_{ok}$ corresponds to the current value corresponding to the diameter considered optimal by the regulations and $F_{ok}$ is the current value for which the value of $F_m$ considered optimal is reached, on average. The lines corresponding to both frequencies of all these significant values were represented. $F_{ok}$ and $D_{proy}$ coincide for both frequencies, in the case of $D_{min}$ and $D_{ok}$, and is slightly higher in both cases for 7 kHz.

Two stages in the evolution of spot diameter with welding current can be clearly distinguished: growth and stabilization. The almost linear relationship between the two parameters during the initial period has been reflected by numerous authors [21,25,50]. During this period, nugget growth occurs as more energy is supplied to the joint. As can be seen, the maximum diameter reached is independent of the switching frequency, but there are notable differences in the extent of the first stage. The figure below depicts the area of the inflection points (8.2 KA and 9.0 kA for 1 KHz and 7 kHz, respectively). As the welding nugget provides the mechanical bond between the two steel sheets, it is to be expected that the mechanical resistance achieved with both switching frequencies is the same, an aspect that will be addressed later.

For both frequencies the loss of stability occurs for current values of 9.6 kA, a value from which projections already occur.

From current of 7.2 kA to current of 8.8 kA, range that includes $F_{ok}$ and $D_{ok}$ and therefore the welding range where the optimal lobe must be considered, the diameter is always greater for switching frequency of 1 kHz. In this range the variability of the diameter is greater for switching frequency of 7 kHz, this is given not only by a decrease in the size of the diameter but also by a greater deviation. A value to note is current of 9 kA where the diameter size of 7 kHz exceeds 1 kHz and the deviation of the latter is greater than the former. For values lower than $D_{min}$, that is, not valid at the quality level, the diameter for 7 kHz is always greater. The same happens for the other invalid range, from $D_{proy}$ where the diameter of 1 kHz also suffers a marked decrease.

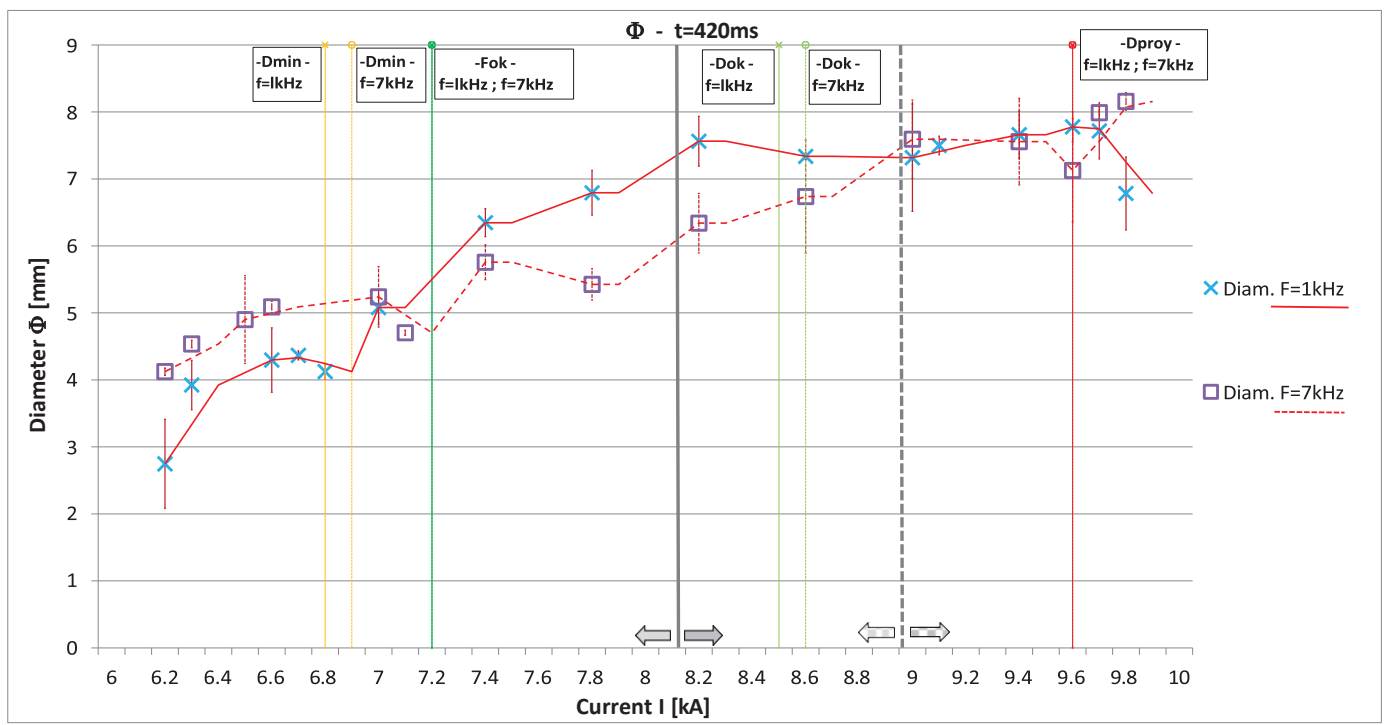

**Figure 11.** Nugget diameter corresponding to 1 kHz and 7 kHz frequencies. Limit values $D_{min}$ and $D_{proy}$ and optimal $F_{ok}$ and $D_{ok}$ for both frequencies.

According to [25], which compares AC and MFDC equipment, the MFDC process produces larger diameter welds than AC at the same current. In the work of Li et al., it is also established that the diameter difference is larger for low currents, and very small when approaching the projection zone. As can be seen in Figure 11, the experimental results obtained in this work are consistent with this trend only in the region of the stabilization values. However, for current values above 7 kA the nugget diameter is smaller at 7 kHz. In the opinion of the authors, the better performance of the MFDC with respect to AC could be attributed to a combination of factors, including the higher frequency of the MFDC, but the effect of current rectification on the results obtained must also be considered. The relative position of the diameter–current curves can be explained by taking into account the nugget rate, which is lower as the excitation frequency increases.

*3.5. Hardness Analysis*

Table 5 shows the hardness results corresponding to the extreme values of the weldability lobe. A clear trend of greater hardness was observed at the switching frequency of 1 kHz.

Figure 12 represents the longitudinal hardness profiles of the indicated significant specimens. The left column corresponds to values at 1000 Hz, from top to bottom: $D_{min}$ and $D_{Proy}$; the right column is the equivalent for 7000 Hz. The vertical lines represent the boundaries of the heat-affected zone and the melting zone, and, between the red lines, is the FZ. Between the green and red lines is the HAZ. It is important to note that, except for the 7 kHz switching frequency and onset of projections, and in contrast to the behavior of DP steels [20], there is no softening at the onset of HAZ. In the case of DP steels, the higher the strength of the steel, the greater the decrease in hardness in the HAZ [8], so a marked softening effect could be expected in CP1000 steel, which has a very high mechanical strength.

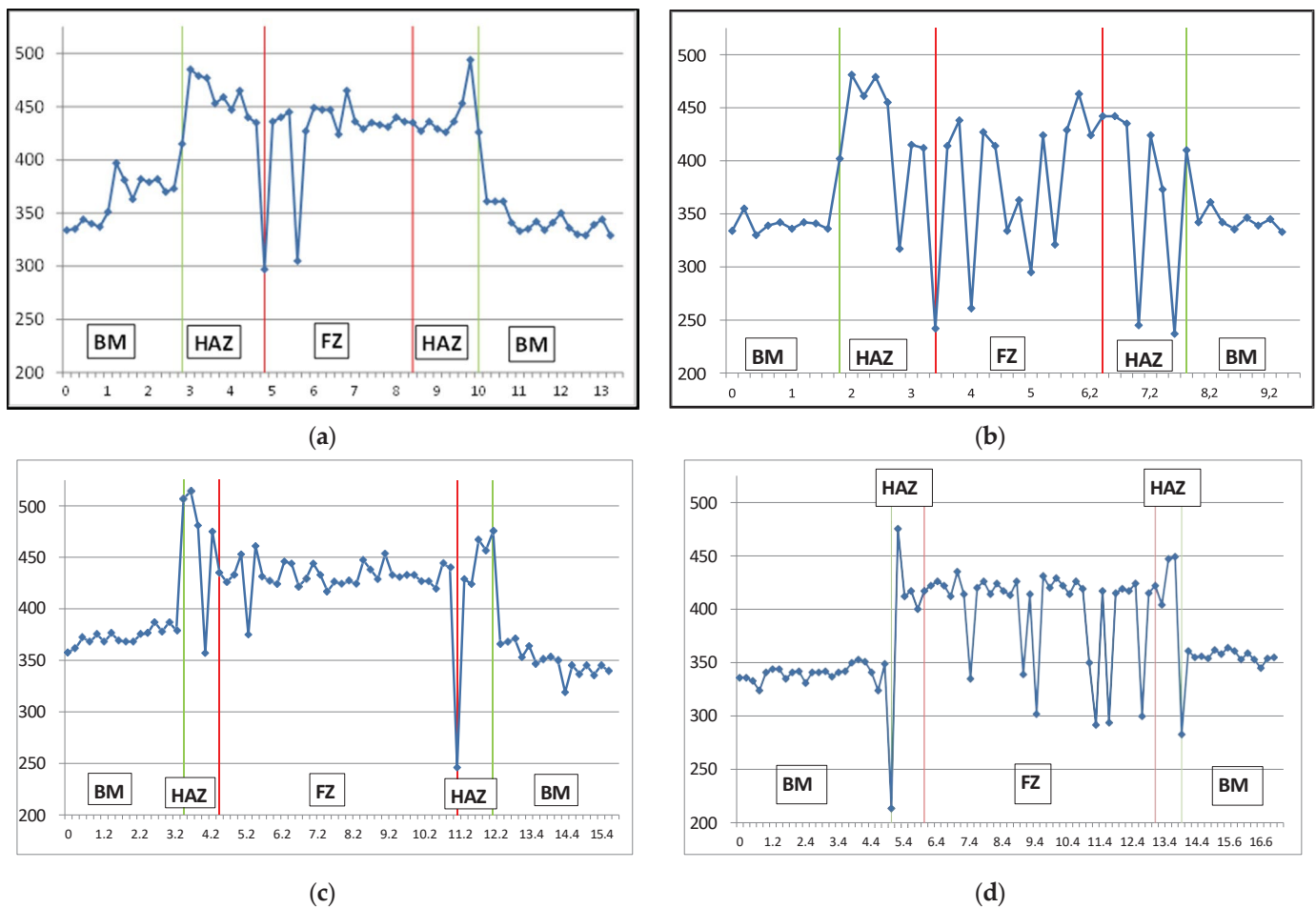

**Figure 12.** Longitudinal line HV t = 420 ms: (**a**) $D_{min}$ f = 1000 Hz, (**c**) $D_{Proy}$ f = 1000 Hz; (**b**) $D_{min}$ f = 7000 Hz, (**d**) $D_{Proy}$ f = 7000 Hz.

In the opinion of the authors, the different behavior between DP and CP steels could be explained as follows: DP steels present a ferritic–martensitic microstructure and the softening in the HAZ is due to the tempering of martensite being the effect more pronounced in the higher mechanical strength grades, since these present a higher percentage of this phase [8].

On the other hand, CP steels are constituted by a bainitic matrix with ferrite and martensite. Due to their chemical composition, the formation of bainite is favored during quenching; in addition, the low hardenability of austenite with a high value of the onset temperature of transformation to martensite, Ms, leads to the formation of auto-tempered martensite [51]. Since this auto-tempered martensite is present in the base metal no softening effect would be appreciated in the HAZ.

Another aspect of great interest is the large hardness variability obtained in the HAZ and FZ zones. Although this effect has already been observed in other works [18,21], attributing it to the rapid heating and cooling during welding that results in phase transformation, the variation of the hardness values obtained in this work is much greater. To explain this difference in behavior, it is worth remembering that the hardness profiles obtained in this study were carried out with a load of 0.1 Kgf, selected instead of the usual 0.2 Kgf, in order to obtain much smaller indentations and, therefore, provide useful information regarding the hardness of the different structural constituents. The low hardness values, between 250 HV and 300 HV would therefore be related to the presence of ferrite, which was more abundant when using the 7 kHz switching frequency.

Some authors [52,53] have studied the effect of the application of an external magnetic field on the weld quality of the RSW, so its use when high switching frequencies are used in the welding of complex phase steels is of interest for future research work. As a future case study, the welding cycle can be modified and the relationship with longitudinal hardness can be found.

*3.6. Shear Tensile Test*

In industry, the use of the tensile test to evaluate the quality of a welded product is common. Each product and each company has its requirements established according to the regulations, needs and final application of the element. The quality departments make use of different methods and tests, being the tensile test the one that serves on many occasions to confirm the quality objectives achieved, it is therefore, despite being a destructive test, one of the test methods widely used. This test makes it possible to compare the efficiency of the welded joint in comparison with the base metal used—not infrequently it is established as a criterion of a resistance weld point; the debonding of this aspect, button failure, that in most of the elements is not a necessary requirement, in addition, as has been mentioned, for the AHSS to achieve this is an added difficulty. The search for a debonding implies an increase in energy to carry out the welding a higher cost, it can also lead to less stabile results due to the proximity of the projection zone. It should be considered that the failure mode is influenced by the type of test, being easier to obtain with a chisel test than with a shear strength. Some authors [20] have focused their study on the transition from one failure mode to another based on the increase in weld diameter, among other aspects. In the experimental results it has been possible to verify that, for CP1000 G10/10 steel, the increase in diameter does not imply a change to the type of button failure in the shear strength, in fact, this type of failure has only been found in a specific way. Debonding has been achieved with greater probability in the chisel test cases, than for the same shear strength parameters with interfacial failure result. Therefore, the shear tensile test to be used according to ISO-14273 [40] is considered more restrictive and repetitive. Figure 13 shows the points corresponding to the mean values of the welds made for each current, their deviations and the trend lines corresponding to the $F_m$ for 1 kHz and 7 kHz.

As discussed above, the vertical lines correspond to the significant values of the weldability lobe. The value of $F_{ok}$ represented in Figure 13 is reached at a welding current of 7.2 kA for both frequencies. The tests carried out verify this value of $F_{ok}$, since, for a switching frequency of 1 kHz, at a welding current of 7 kA there is an $F_m$ = 15.82 kN, considering that the $F_{ok}$ has been established at $F_{mok}$ = 15.83 kN, this value will be achievable around the value indicated in Figure 13, that is, for a welding current of 7.2 kA. With respect to the switching frequency of 7 kHz, the variability of the values of $F_m$ makes it reach the value of $F_{ok}$ at a welding current of 6.6 kA, but it falls below that value for a current of 7 kA, reaching it again for a current of 7.1 kA, which indicates that the stabilized $F_{ok}$ value occurs around the value in Figure 13, welding current of 7.2 kA. The maximum $F_m$ value for both frequencies is reached around current of 9.4 kA, approaching the projection zone.

It is observed that the establishment of the $F_{ok}$ criterion leads to the welding current necessary to achieve the desired quality being considerably less than that necessary to achieve the $D_{ok}$ value, for either of the two frequencies. It is observed that the central area of the lobe, between welding current of 7.2 kA and 8.8 kA, is characterized by the $F_m$ of 1 kHz always above the corresponding 7 kHz, and the deviation in the same range of 7 kHz is higher. If the current is increased outside this range, something that is not advised, since the advantage obtained is not significant, as the $F_m$ practically does not increase, but, if its deviation does, it is observed that the deviation from switching frequency of 1 kHz increases reaching values that make it non-functional past the projection line. It is interesting to highlight this aspect because an incorrect parameterization of the equipment is more problematic for 1 kHz, while if the parameterization is correct, the advantage of switching frequency of 1 kHz is clear.

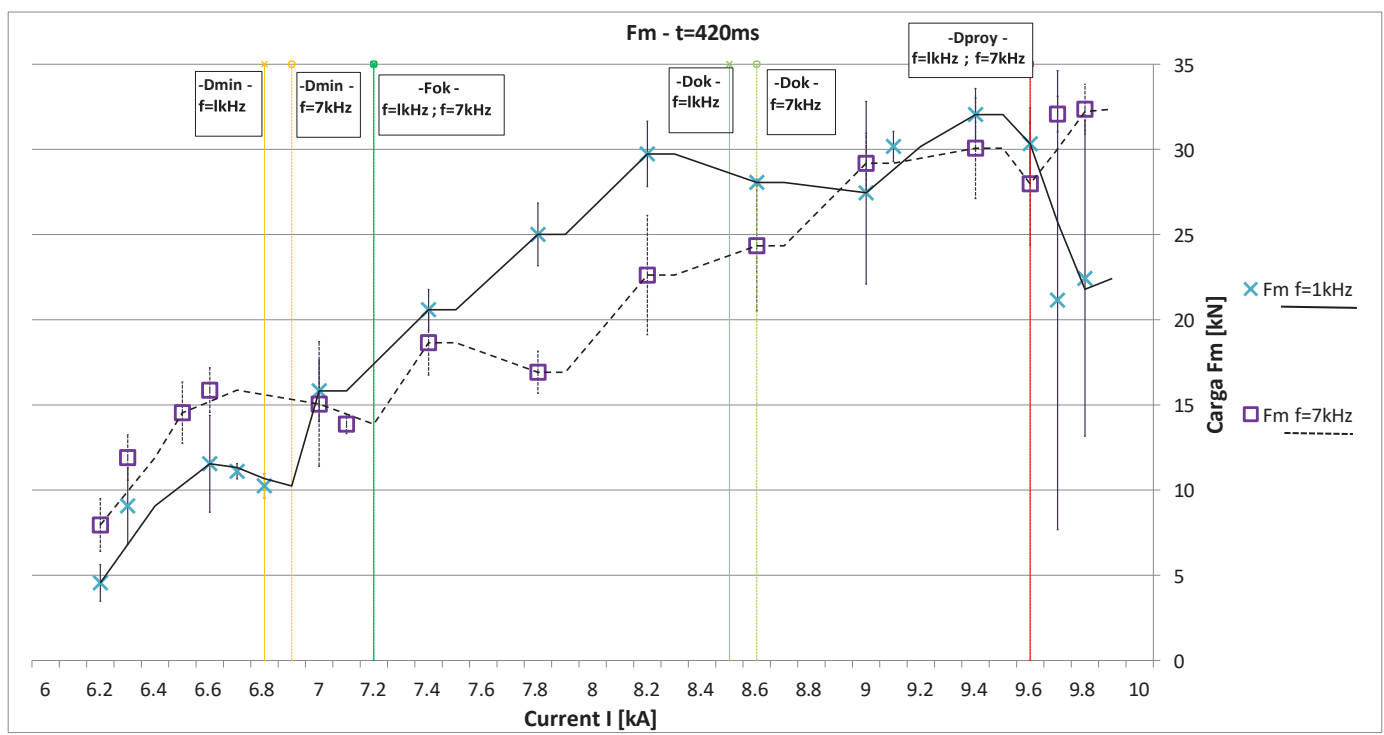

**Figure 13.** Shear tensile test for the study frequencies, 1 kHz and 7 kHz. Limit values $D_{min}$ and $D_{proy}$ and optimal $F_{ok}$ and $D_{ok}$ for both frequencies.

As can be seen in Table 6, there are cases in which despite having a higher Φ the $F_m$ is lower, this can be observed at values at a switching frequency of 7 kHz, between values of 1 kHz and 7 kHz and between values of 1 kHz and 7 kHz, after removing the zinc coating layer. In losing this direct relationship between Φ and $F_m$, it was decided that $F_m$ be taken, directly, as a criterion in the weldability lobe and in the quality analysis.

**Table 6.** Justification of the $F_m$ criterion for the weldability lobe.

| Frequency | | | f = 1000 Hz | | f = 7000 Hz | |
|---|---|---|---|---|---|---|
| F(daN) | T(ms) | I(kA) | φ(mm) | $F_m$(kN) | φ(mm) | $F_m$(kN) |
| 470 | 210 | 11 | 5.43 | 16.95 | 5.39 | 17.22 |
| 470 | 260 | 10 | 5.65 | 17.51 | 5.64 | 18.01 |
| 470 | 420 | 7 | 5.08 | 15.83 | 5.24 | 15.05 |
| 572 | 210 | 10 | 4.69 | 12.61 | 4.76 | 12.57 |
| 572 | 210 | 11 | 5.27 | 16.52 | 5.49 | 16.41 |
| Galvanized chemically removed | | | | | | |
| 470 | 210 | 10 | 6.70 | 25.69 | 6.80 | 25.45 |

### 3.7. Efficiency

The efficiency of an industrial process is of great importance. The establishment of a process as efficient and functional can be carried out thanks to efficiency studies, which is why, together with the criterion of the load force $F_m$, the efficiency η is the other aspect established as a quality criterion in this study. The welding process must meet the quality criteria based on the established $F_m$ requirements, but it must also later prove to be an effective process capable of producing benefits and therefore efficiently adoptable by real industry.

Figure 14 shows the mean efficiency values of the welds performed for each electric current, their deviations and the efficiency trend lines for the two study frequencies. The efficiency of the total process has been represented in green, that is, from the three-phase input to the welding point in the specimens. The efficiency corresponding to the converter is represented in blue, that is, from the three-phase input to the primary input of the transformer. The yellow color represents the efficiency corresponding to the stage of the transformer + rectifier + interconnections, that is, from the primary of the transformer to the welding point in the test specimens.

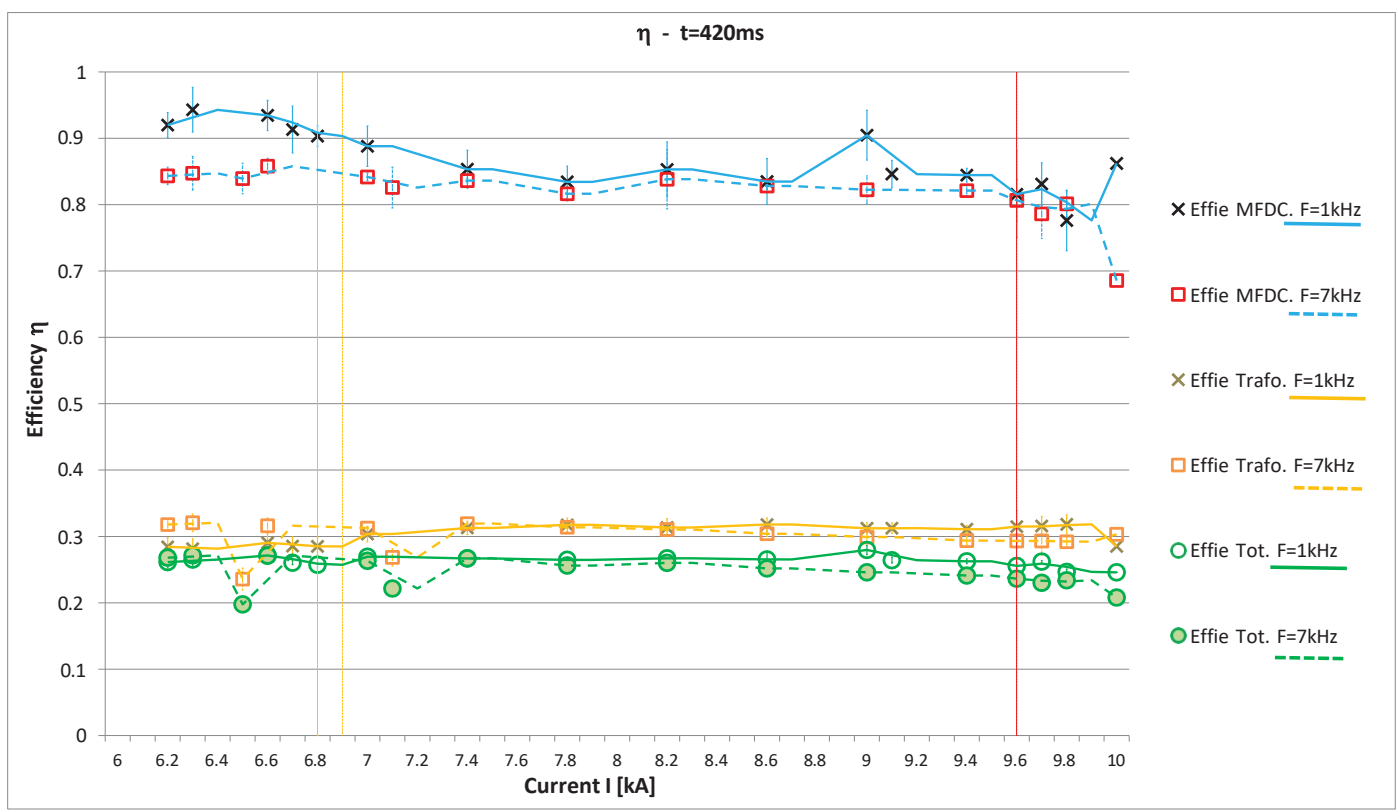

**Figure 14.** Converter efficiency for 1 kHz and 7 kHz in blue; efficiency of the transformer + rectifier + interconnection block for both frequencies in yellow; efficiency of the complete process for the two study frequencies in green.

The highest efficiency is in the block corresponding to the converter, being slightly better for a switching frequency of 1 kHz, which verifies that the switching losses in semiconductors are greater for a higher frequency. The efficiency corresponding to the transformer + rectifier + interconnections set is considerably lower than that of the converter, which makes this block the least efficient, due to the losses in the transformer, in the rectifier and in the final wiring. For this second block, the efficiency at medium currents does not present large differences for the frequencies studied, it wass found that the efficiency was slightly better for 1 kHz. This agrees with [29] who state that a switching efficiency improves as an inverse function of frequency. The total efficiency was higher for 1 kHz, from the valid current values, to obtain a quality weld. The total efficiency dropped as the current is increased, this effect being greater for 7 kHz. The low deviation in all values is noteworthy, being slightly higher with the converter working at 1 kHz.

Comparing Figure 14 where the efficiency trend lines are shown against Figure 13 where the $F_m$ loads are shown, it is observed that the total efficiency remains at high values for the $F_{ok}$ value, while as one moves towards at $D_{ok}$ the efficiency value falls for switching frequency of 7 kHz. Which reinforces the fact of using a $F_{ok}$ value instead of the $D_{ok}$ with lower efficiency, thus avoiding in many cases an unnecessary cost.

## 4. Conclusions

The study here reported has shown that the operating frequency of the inverter used in the RSW equipment has a considerable influence on the quality of the welding process. For this, 481 test specimens were made; the diameter of 419 weld points was measured; 339 shear strength tests were performed; 52 hardness tests; 53 welding points were analyzed macrographically, and 16 through SEM. A total of 1360 tests were carried out on the welded specimens.

On the one hand, it has been shown that the quality of the welded joint, in terms of mechanical strength, absence of defects, efficiency of the welding process, is better when the lower frequency is used.

On the other hand, this work showed that the main advantage of using higher frequencies is that the parameterization of the welding equipment is easier, due to a wider weldability lobe.

It should be noted that a great variability was found in the hardness profiles, greatest in the case of 7 kHz. The presence of surface cracks was also greater for 7 kHz, although the size of the internal pores was smaller.

The expulsion of zinc improved with increasing welding current and with the switching frequency.

The $F_m$ is clearly better for 1 kHz within the values of the weldability lobe.

Most of the losses occur in the output stage: transformer + rectifier + interconnections, being greater with increasing frequency and current.

**Author Contributions:** Conceptualization, G.M.-S.; methodology, G.M.-S., A.C. and J.D.-G.; validation, G.M.-S., A.C. and J.D.-G.; formal analysis, A.C. and J.D.-G.; investigation, G.M.-S.; resources, G.M.-S., A.C. and J.D.-G.; data curation, G.M.-S.; writing—original draft preparation, G.M.-S.; writing—review and editing, G.M.-S., A.C. and J.D.-G.; visualization, G.M.-S.; supervision, A.C. and J.D.-G. All authors have read and agreed to the published version of the manuscript.

**Funding:** This work was supported in part by the Spanish State Research Agency (AEI) under project PID2019-105612RB-I00/AEI/10.13039/501100011033 and in part by the Government of Galicia under the project GPC-ED431B 2020/03.

**Institutional Review Board Statement:** Not applicable.

**Informed Consent Statement:** Not applicable.

**Acknowledgments:** This research has been possible thanks to the donation of the materials used for the experiments by GAESA-Ipm Rubi Group, and the structure of the equipment donated by Inoxidables Fegosan.

**Conflicts of Interest:** The authors declare no conflict of interest.